# Algorithmic Authority: The Case of Bitcoin

Caitlin Lustig
University of California, Irvine
clustig@uci.edu

**Abstract**

*In this paper, I propose a new concept for understanding the role of algorithms in daily life: algorithmic authority. Algorithmic authority is the legitimate power of algorithms to direct human action and to impact which information is considered true. I use this concept to examine the culture of users of Bitcoin, a crypto-currency and payment platform. Through Bitcoin, I explore what it means to trust in algorithmic authority. My study of the Bitcoin community utilizes interview and survey data. I found that Bitcoin users prefer algorithmic authority to the authority of conventional institutions which they see as untrustworthy. However, I argue that Bitcoin users do not have blind faith in algorithms; rather, they acknowledge the need for mediating algorithmic authority with human judgment. I examine the tension between members of the Bitcoin community who would prefer to integrate Bitcoin with existing institutions and those who would prefer to resist integration.*

## 1. Introduction

Algorithms have always been a crucial part of software development—they are "the fundamental entity with which computer scientists operate" (Goffey, 2008). Beyond just shaping the behavior of software, algorithms also play an increasingly critical role in shaping societal and individual behavior.

David Beer (2009) argued that with the Web 2.0, software became ubiquitous and participatory, which gave algorithms "the capacity to shape social and cultural formations and impact directly on individual lives". An example of the increasing power of algorithms in everyday life is Google's suite of search and ranking algorithms, which influence the information users have access to and in turn may impact what they judge to be true (Introna and Nissenbaum, 2000). Gillespie (2014) refers to these kinds of algorithms as "public relevance algorithms"—algorithms that "select what is most relevant from a corpus of data composed of traces of our activities, preferences, and expressions".

Algorithms not only impact how users constitute their knowledge of the world, but they also can act as employers or managers of users. This role of algorithms in shaping human behavior can be observed in technologies of *heteromation*, which are characterized by their use of humans as integral "computational components" to augment software algorithms. Heteromation is a reversal of automation, which uses computers to perform prohibitively expensive or difficult tasks for humans. Instead, heteromated technologies use humans to perform tasks for computers (Ekbia and Nardi, 2014). The reasons for using heteromation are varied—in some cases, it is simply cheaper and easier to have humans perform some computational tasks. In other cases, the use of heteromation may be more strategic—the affective rewards of heteromation may induce brand loyalty by making users feel useful and involved in a product. Heteromated systems usually have beneficiaries that generally receive large financial benefits from the unpaid (or underpaid) work of heteromated laborers. Heteromated laborers are generally "paid" in affective rewards. One way in which heteromated systems differ from other systems in which laborers are underpaid is that heteromation uses computational systems to organize and manage laborers, and they run computations on the work that the laborers perform. Although the type of work that the concept of heteromation encompasses is quite broad, heteromated systems typically involve human work as a necessary part of the algorithms they use.

An example of heteromation is Amazon Mechanical Turk (AMT), in which "humans [are] rendered as bits of algorithmic function" (ibid). AMT workers are generally recruited to perform tasks that cannot easily be performed by artificial intelligence. In fact, Amazon has coined the term "artificial artificial intelligence" to refer to the ways in which "computer scientists […] work on integrating the human work made computationally accessible by Amazon into existing data systems, artificial intelligence training algorithms, and interactive applications" (Irani, 2013). Requesters also use algorithms to manage workers, a sort of "automated management", by determining which tasks to show workers and to help them select tasks they would be best suited for (ibid).

I call the emergent dynamic between algorithms and human actors "algorithmic authority". Algorithmic authority is defined as *the authority of algorithms to direct human action and to verify information, in place of relying exclusively on human authority*. In the case of AMT, the heteromation of work necessitates that workers are under control of an algorithmic authority. The primary role of this authority is to select workers to perform certain tasks. The employers trust AMT to produce good results because the filtering mechanisms for selecting works are automatic and sophisticated. However, as I demonstrate in this paper through a study of Bitcoin, *algorithms appear to have authority because they are automatic and self-contained, but in actuality, users must mediate algorithmic operations*.

For AMT requesters, the algorithm obscures the hybrid system of humans and code that are performing labor. Although requesters are aware that they are employing humans, AMT strips away much of the human aspect of this work by rendering humans as nameless computation components with which requesters do not need to interact with directly. "Sociotechnical assemblages black box the complex politics of management into familiar acts of writing code and manipulating spreadsheets. By rendering the requisition of labor technical and infrastructural, the design of AMT limits the visibility of workers, rendering them as a tool to be employed by the intentional and expressive hand of the programmer" (ibid). AMT requesters must trust both the filtering algorithm and their workers, even if they are rendered largely invisible by the code. A corollary, then, to the definition of algorithmic authority is that *algorithms have authority because users trust both the algorithm and the assemblage of sociotechnical actors that mediate the algorithm*.

While my definition of algorithmic authority is unique, I distinguish my use of the phrase algorithmic authority from Clay Shirky's. In a blog post, Shirky (2009) described algorithmic authority as "the decision to regard as authoritative an unmanaged process of extracting value from diverse, untrustworthy sources, without any human standing beside the result saying 'Trust this because you trust me.'". Shirky uses Google's PageRank algorithm[1] as an example of how information can be generated automatically and trusted by most people as "legitimate". My use of the term algorithmic authority is broader. I argue that algorithms are given authority to not only decide which information is true but also to direct human action within heteromated systems.

In this paper, I examine what kind of trust people put in algorithms, what it means to trust in algorithms, and how users mediate algorithmic authority with their own personal judgment and the judgments of other, trusted people. I examine these questions through a qualitative study of Bitcoin, a crypto-currency that on October 15, 2014 had an exchange rate of $393.12 USD to one bitcoin[2]. I chose Bitcoin as an example of algorithmic authority because it is not managed by governments or banks, but by algorithms. The currency is also notably different from many public relevance algorithms, which are perhaps the most discussed instances of algorithmic authority. Nothing in the Bitcoin algorithms inform users of how to act or are in any way customized to the user. Furthermore, unlike Google's search engine optimization algorithms, which would be impossible to fully understand and predict even if the code was open source[3], Bitcoin's code is open source and its users often openly discuss its algorithms. They are so often discussed that Maurer et al. argue that the code is one of two things[4] that are "*practically all that Bitcoin enthusiasts ever talk about*" (2013). Indeed, my participants were well informed about the ways in which Bitcoin's code worked and frequently thought about ways to improve the algorithms that Bitcoin employs.

The main Bitcoin algorithm that I will refer to is the *blockchain*, a mechanism for producing the currency and verifying transactions. In this paper, I follow the convention of Bitcoin users by referring to the system as Bitcoin and the units of currency as bitcoin.

## 1.1. Algorithms

In order to explain the concept of algorithmic authority, I must first define the two parts of the phrase. The most basic definition of algorithms is that they are "logic + control" (Kowalski, 1979). They are simultaneously a set of abstract instructions (logic) and possibilities for action (control). Goffey (2008) suggested, "One of the implications of characterizing the algorithm as a sum of logic and control is that it is suggestive of a link between algorithms and action. [….] Algorithms do things, and their syntax embodies a command structure to enable this to happen". Algorithms that are given authority not only "do things" to software, but they also can cause human actors to respond accordingly, both indirectly (through the knowledge construction of public relevance algorithms) and directly (through algorithms of heteromation).

---

[1] In actuality, Google has a number of search and ranking algorithms, but PageRank is the most frequently referenced.

[2] https://coinbase.com/charts
[3] Ziewitz suggests, "Even if you had Larry and Sergey at this table, they would not be able to give you a recipe for how a specific search results page comes about" (2012).
[4] Maurer et al. argue that Bitcoin enthusiasts also talk about the labor involved in mining bitcoin.

In their article, "Governing Algorithms: A Provocation Piece", Barocas et al. (2013) question the popularity of the word "algorithms" in academic research on sociotechnical assemblages. They ask, "Would the meaning of the text change if one substituted the word "algorithm" with "computer", "software", "machine", or even "god"? What specifically about algorithms causes people to attribute all kinds of effects to them?" I would further ask, if algorithms are the building blocks of software—what is the discursive advantage to using "algorithms" as our unit of scale?

The reason I use the word "algorithm" is that it invokes "what many see as computing's distinctive features: a combination of rigid procedurality, formalism, and quantification—the abstract concreteness of computation" (Seaver, 2013). Much of the literature on the social impacts of algorithms seems to critique this procedurality, arguing that large-scale algorithms are inscrutable black boxes which have great power over our daily lives, but we cannot know exactly how they come to the decisions that they do. Consequently, we have little recourse when algorithms discriminate against us (Pasquale, 2011). Algorithms are seen as frighteningly uncaring when it comes to making decisions that have a very real impact on human lives.

When I use the word "algorithm", I use it to join in on this conversation about the values of algorithms. Barocas et al. note that this conversation suggests that algorithms have a certain agency, that algorithms "do things", which of course, Goffey's definition of algorithm does argue. Barocas et al. criticize this idea that algorithms have agency because it obscures the responsibility of the people who developed the algorithm by suggesting that algorithms can act on their own. In this provocation, Barocas et al., do not mention responsibility of users who input data into the algorithm, but nevertheless, their point stands—it is not enough to simply place all of the responsibility on the code. The agency and action of algorithms neither resides entirely in the code nor entirely in the human actors who use or develop the code. Furthermore, the ability to do things is not evenly distributed among all actors—developers can change the algorithm in ways that users cannot, and algorithms may restrict the actions of their users.

In this paper, when I discuss an algorithm "doing something", I am not claiming that the action purely lies within the code. In fact, Gillespie (2014) reminds us that algorithms can be done by hand. Rather, I will develop the notion that algorithms with authority over users are best analyzed as part of sociotechnical assemblages. These assemblages consist of not only the code, but also a number of human actors, such as programmers, users, and regulators. While it may be the case that algorithms do things, they do things because they were programmed to do them, because users trust them to take certain actions, and because regulators allow them to do things.

## 1.2. Authority

For my definition of authority, I draw from Max Weber. Weber (1922) described authority as leadership that is perceived as legitimate and without coercion. Weber defined three kinds of authority: traditional, charismatic, and rational-legal authority. Traditional authority refers to authority that derives its legitimacy from precedents and social norms. Charismatic authority refers to an authority that derives legitimacy from a "higher power" and challenges traditional authority. Rational-legal authority refers to "appeal to efficiency, and the rational fit between means and intended goals" (Coleman, 2013).

If we take Weber's taxonomy of authority as true, which of these forms of authority do algorithms utilize? Although the authority of algorithms is in no way limited to rational-legal authority, it seems to draw most heavily from those ideals. (Indeed, the religious language used to describe Bitcoin's creator points to some kind of charismatic authority by proxy for the currency.) An algorithm must be represented as objective in order to gain legitimacy through rational-legal authority. In addition, Gillespie (2014) argues that in order for public relevance algorithms to be seen as legitimate, they must appear to be completely automated. In order to maintain the legitimacy of the "shape" of the algorithm, both technically and in terms of the values it represents, it must be described by the developers of the algorithm in a way that defends the impartiality of the algorithm. However, Gillespie points out that algorithms are often not completely objective and, furthermore, they are often not completely automated or "hands off". There are many things that Google's search algorithms may find relevant to a query that they simply will not display for legal or moral reasons. For example, Google "refuse[s] to autocomplete search queries that specify torrent file-trading services". (ibid) Another example, was how in 2010, PayPal decided to freeze donations to accounts associated with the controversial website WikiLeaks (Maurer et al., 2013).[5]

### 1.3. Trust

Trust is of particular importance in understanding how algorithms gain authority. I turn to Gambetta's definition of trust to understand what it means to trust in algorithmic authority:

> trust (or, symmetrically, distrust) is a particular level of the subjective probability with which an agent assesses that another agent or group of agents will perform a particular action, both *before* he can monitor such action (or independently of his capacity ever to be able to monitor it) *and* in a context in which it affects *his own* action. (1990)

Algorithms generally have desired outcomes, such that savvy users can predict the algorithm's outcome. They can trust that the software will behave in a certain way and act accordingly in response. Thrift (2005) stated that "software is best thought of as […] an expectation of what will turn up in the everyday world […] a means of sustaining presence which we cannot access but which clearly has effects, a technical substrate of unconscious meaning and activity" (p. 156). Software can shape our expectations and understanding of the world, but Thrift argued that while its effects are visible, the relationship between the inputs to the software and the exact mechanisms used are largely invisible, creating a sort of "technological unconscious". Users may trust in software because of some level of predictability, but they are vulnerable to the invisible effects of software.

## 2. Background

Bitcoin was developed in 2008 during what has come to be known as the Great Recession. This financial crisis led to significantly higher rates of unemployment as well as lower wages for those who were still employed (Blendon and Benson, 2009). Overall, Americans were pessimistic about stock market and housing prices (Hurd and Rohwedder, 2010). In 2009, 48% of Americans reported that they felt angry that the government was "[b]ailing out banks and financial institutions that made poor financial decisions"[6]. During and subsequent to the Great Recession, growing discontent with governments and capitalism led to the popularization of protest movements such as the Occupy movement (Trudell, 2012). Occupy protesters argued that the current way in which liberal democracies function is deficient (Razsa, M. and Kurnik, 2012).

Like members of the Occupy Movement, many Bitcoin users are drawn to the currency because they wish to disrupt institutions by creating their own institutions governed by consensus. To understand what it means to "disrupt" an institution, I turn to Schumpeter's (1942) definition of "creative destruction", i.e., that which "incessantly revolutionizes the economic structure from within, incessantly destroying the old one, incessantly creating a new one". Bitcoin is appealing to users because it not only provides an alternative to government-based fiat currency[7], but because they believe it can transform governments and economic systems.

Bitcoin was first introduced in the whitepaper "Bitcoin: A Peer-to-Peer Electronic Cash System" by Satoshi Nakamoto (2008). It is widely believed that Satoshi Nakamoto is a pseudonym for the person or persons who created Bitcoin. Nakamoto's motivation for creating a new type of currency was based on the shortcomings of electronic commerce using conventional currencies. In his seminal paper on Bitcoin, Nakamoto argued that electronic commerce is flawed because it has high transaction fees and affords consumers less privacy than cash transactions.

---

[5] There was considerable political pressure for financial institutions to put a blockade on WikiLeaks. WikiLeaks was still able to partially get around this blockade by accepting bitcoin donations (Matonis, 2012). Although Satoshi Nakamoto, the creator of Bitcoin, was opposed to being associated with WikiLeaks, because Bitcoin has no centralized authority, it would have been impossible to bar WikiLeaks from receiving bitcoin donations. He said, "I make this appeal to WikiLeaks not to try to use Bitcoin. Bitcoin is a small beta community in its infancy. You would not stand to get more than pocket change, and the heat you would bring would likely destroy us at this stage" (https://bitcointalk.org/index.php?topic=1735.msg26999#msg26999, 2010).

[6] http://www.pewresearch.org/daily-number/bailing-out-banks/
[7] Fiat currency refers to currency whose value is not derived from a commodity, but from an authority (usually a government). Some debates remains as to whether Bitcoin is a fiat currency, because while it is not a government-issued currency, it *arguably* not a commodity—there is no practical use for Bitcoin besides as a currency. Kaplanov argues that, "[u]nlike fiat currencies, whose value is derived from regulation or law and underwritten by the state, bitcoins have no intrinsic value and the only real value is based on supply and demand—what people are willing to trade for them" (2012). However, Grinberg argues that Bitcoin is fiat currency simply because it is not a form of commodity money (2011). Bitcoin users themselves tend to frame Bitcoin as oppositional to fiat currencies, which suggests that they do not see it as a fiat currency. The confusion is complicated even more by some participants reporting that they *do* use Bitcoin as a commodity. The ways in which Bitcoin is used as a commodity are further explored in section 5.4. On participant, Keith, stated, "I believe Bitcoins are a commodity trying to function like a currency. In reality Bitcoins are more like bearer shares but the Bitcoin community wants it to act like a currency."

According to Nakamoto, the methods taken to prevent *double spending* are the main cause of the shortcomings of electronic commerce. Double spending refers to spending money in one online purchase, and then quickly making another purchase with the same money. To prevent double spending online, trusted third parties must verify transactions for merchants. However, these third parties charge transaction fees, which limit the types of transactions that are feasible by making micro-transactions prohibitively expensive. Third parties can also reverse transactions when there is customer fraud. While a third party's ability to reverse transactions protects customers from identity theft, it also puts merchants at risk of losing money. In order to gain some level of trust in their customers, merchants must acquire information about their customers to confirm their identity. Consequently, consumers are unable to conduct anonymous or private transactions. Nakamoto (2008) argues that, "[t]hese costs and payment uncertainties can be avoided in person by using physical currency, but no mechanism exists to make payments over a communications channel without a trusted party". In response to these issues, Nakamoto designed Bitcoin to solve the issue of double spending without mandatory transaction fees or loss of privacy.

Nakamoto created Bitcoin as a currency and payment platform that is *pseudo-anonymous, supports micro-transactions,* and *has no inherent transaction fees*. Bitcoin has the same affordances as cash, but in a digital format. While cash typically relies on a government to regulate it, Nakamoto's system relies on a distributed and decentralized user base to give it value through consensus and artificial scarcity—two concepts that are ingeniously tied together in Bitcoin's implementation. The user base forms a peer-to-peer network that keeps a public, pseudo-anonymous ledger of all transactions. This ledger is called the *blockchain*. When a transaction is made, the Bitcoin software running in the background on peers' computers uses calculations to verify that the money has not been double spent by comparing it against this ledger. In order to incentivize users to run this software, there is a random chance that the user will be rewarded new bitcoins for completing a calculation. Bitcoin users refer to the generation of new bitcoins as "mining" bitcoins. When bitcoins are mined, the verified transactions are put into the blockchain.

The blockchain mechanism has been described as "the main innovation of Bitcoin"[4]. Through the blockchain, verified transactions are bundled into blocks that are made a part of the public ledger. These blocks are connected to each other through the use of hash algorithms, which cryptographically generate a unique value to represent each block. Every block contains the hash of the previous block in order to indicate its position in the blockchain. In the case that two or more blocks are mined within a few seconds of each other, the Bitcoin client software may be unable to determine which block should come next in the blockchain. In this situation, the blockchain creates a fork. Whichever branch of the fork is longer becomes a part of the main blockchain and the other branches are discarded. In Figure 1, the longest chain is shown in black and the other branches are unused.

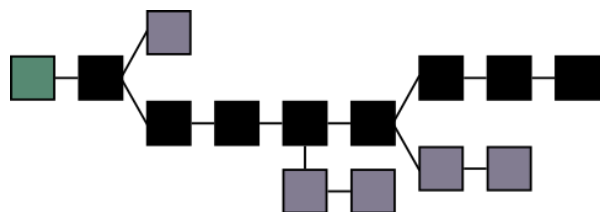

**Figure 1: A graphical representation of the blockchain mechanism.**[8]

The mining process serves as a way to control the growth rate of Bitcoin and eliminates the need for a centralized agency to distribute the currency. The code of Bitcoin dictates that there are 21 million bitcoins that can ever exist. As of June 15, 2014 nearly 13 million bitcoins had been mined. Bitcoin is designed such that on average, a block of bitcoins will be mined every ten minutes. Every four years, the number of bitcoins that are mined in each block goes down by 50%. Currently, 25 bitcoins are mined in each block. In 2017, the number of bitcoins in a block will be reduced to 12.5. Users have calculated that by the year 2140 all Bitcoins will have been mined. Most users who mine bitcoins pool together their resources and join a "mining pool". Every time the pool mines bitcoins, the coins are distributed in proportion to the amount of processing power that the miner contributed.

Bitcoin is hardly the first modern widespread alternative to fiat money, but it is the first to capture the public's attention. Bitcoin has been compared to two previous alternative currencies, Liberty Dollars (1998-2009) and e-gold (1996-2009). Liberty Dollars was shut down because the currency was in direct competition with and closely resembled the US dollar (Grinberg, 2011). E-gold was shut down due to the corporation's lack of licensing as a "money transmitting business" and its inconsistently enforced policies against illegal activity. Unlike these currencies, Bitcoin is commonly characterized as being in a legal gray area. Because of Bitcoin's decentralized nature, it may be harder to shut down, as there is no central organization to prosecute. Unlike e-gold, there is no infrastructural mechanism to determine the identities of users (Grinberg, 2011; Kaplanov, 2012).

---

[8] https://en.bitcoin.it/wiki/Block_chain

## 3. Related Work

The idea of anonymous digital currencies is hardly new. Some early examples include Chaum et al.'s (1990) untraceable cash and NetCash (Medvinsky and Neuman, 1993). These systems differed from Bitcoin in that they were not decentralized. Chaum's idea relied on the cooperation of banks and Medvinsky and Neuman's relied on central servers. Systems that rely on peer-to-peer networks to prevent double spending were introduced by Garcia and Hoepman (2005). The proof-of-work system used in Bitcoin is derived from Back's (2002) hashcash, which was originally conceived as a way to discourage email spam by requiring senders to do computationally complex work before sending an email. Bitcoin also borrows from Dai's (1998) b-money, which required that transactions were publically announced in order to verify them in a decentralized manner. b-money had a similar mining mechanism to Bitcoin and utilized Back's proof-of-work concept. In 2004, Finney proposed a system in which proof-of-work could generate reusable token which could be passed along to different users and generate more tokens themselves. Finney's work is the basis for how Bitcoin integrates mining and the blockchain. But perhaps the most closely related work to Bitcoin is Nick Szabo's (2008) bit gold, which applied Finney's work to a monetary system[9]. Instead of generating reusable tokens, bit gold proposed generating currency using the same mechanism. Bitcoin differs from bitgold by a Byzantine-resistant peer-to-peer system (Miller and LaViola, 2014). This difference meant that Bitcoin has certain fail-safe methods built in that should prevent one bad peer from compromising the entire blockchain. To prevent doublespending or other errors in the blockchain from propagating, the majority of the computational power would have to approve a given transaction. For a more detailed description of the ideas that influenced Bitcoin's design, see (DuPont, 2014) and (Barok, 2011). For an understanding of the legal implications of Bitcoin based on previous alternative currencies, see (Grinberg, 2011) and (Kaplanov, 2012).

Crypto-currency is an emergent area of study and consequently there is little research on the Bitcoin user base. As Bitcoin has gained more popularity a number of news articles have been written on the currency, but many focus on illegal activity conducted by Bitcoin users. Few academic articles explore Bitcoin from a social or theoretical perspective. A notable exception to this is Maurer et al.'s research, which examined the Bitcoin user base. They argued that Bitcoin users place their trust in Bitcoin's code to produce and distribute bitcoins correctly, as opposed to trusting a government or a central organization to do so. One reason for this trust is the transparency of Bitcoin's code—users trust the code because of "their collective ability to review, effectively evaluate, and agree as a group to changes to it". Maurer et al. argue that users can trust the code because of "the fact that such decentralization, as well as the public-key encryption of users' identities, is hardwired into the system". Bitcoin is designed to prevent corruption, and consequently, users find the system more trustworthy than institutions such as governments and banks.

Mallard et al. (2014) build on Maurer et al.'s argument and suggest that trust in Bitcoin is distributed through several socio-technical mechanisms, one of which is the underlying algorithms of Bitcoin's code, in particular because it is a peer-to-peer system. To use a peer-to-peer system, users must actively participate by pooling together resources, which builds trust. Users also trust in the resilience of peer-to-peer networks and trust in Bitcoin's core developers.

Other papers that discuss the behavior or ideals of Bitcoin users have largely utilized analyses of the blockchain to learn about the aggregated behavior of Bitcoin users (e.g. Meiklejohn, 2013; Reid and Harrigan 2013; Ron and Shamir, 2013).

## 4. Methods

This study consisted of two phases: a survey and a series of interviews. I also engaged in participant observation by actively reading posts and communicating with users on Bitcoin blogs, forums, and articles. My findings in this study were derived using Grounded Theory methodology (Glaser and Strauss, 2009). I did not begin with the theory of algorithmic authority, rather I developed it after analyzing my data.

---

[9] A common theory about the identity of Satoshi Nakamoto is that he is either Nick Szabo, Wei Dai, or Hal Finney.

### 4.1. Survey

I posted an exploratory survey of 36 questions on BitcoinTalk.org (326,031 users) and /r/Bitcoin, the Bitcoin subreddit (122,561 subscribed users) in October, 2013. I selected these forums due to their popularity and reputation among Bitcoin users. The purpose of the survey was to develop a better understanding of attitudes of the Bitcoin community. The survey was comprised of Likert scale questions to assess study participants' reasons for using Bitcoin; open-ended questions for topics such as anonymity, government regulation and the future of Bitcoin; and multiple-choice demographic questions. At the end of the survey participants had the option of leaving their email address so that they could be later contacted for an interview. Out of the 510 survey participants, 124 left their email addresses.

### 4.2. Interviews

I used the data collected from the survey to craft interview questions based on the themes that emerged from the responses. From the survey data, I noted that participants had diverse views on the future of Bitcoin and government regulation of Bitcoin. I used exploratory and semi-structured interviews to learn more about the diversity of views on those issues, and I iteratively refined the questions based on previous interviews. I contacted a third of the survey participants for interviews in waves until over twenty participants agree to be interviewed. I did not contact survey participants who left short answers or did not answer a large number of questions. All interviewees were male; no women offered to be interviewed. I interviewed 22 participants from March, 2014 to May, 2014. Participants were given the option of being interviewed over any medium they wished: email, instant messaging, Skype, telephone, or in-person (if possible). Eight participants chose some form of voice or video communication, three participants selected instant messaging, and 11 participants chose email.

All participants have been anonymized in this paper; any names used for interview participants are fictitious, with the exception of Frankenmint, a participant who requested that he be identified by his online moniker, which was derived from how he took his "first hosting PC of spare parts to make it and it minted bitcoins with the assistance of miners." Original orthography and punctuation have been preserved for participants who communicated over instant messaging and email.

### 4.3. Participant observation

Over the course of this study, I systematically read Bitcoin forums and blogs from November, 2012 to the present, to immerse myself in the Bitcoin community. I primarily read posts on the website Reddit, a popular social networking and news website which consists of "subreddits"—smaller boards devoted to a specific topic. The forum and blog posts I read focused on the latest Bitcoin news, technical details of Bitcoin's implementation, and political economic theory. I also read mainstream news articles in order to understand the popular perception of Bitcoin users. Prior to the shutdown of the black market Silk Road (which utilized Bitcoin for payment) in October, 2013, news articles tended focus on illicit uses of Bitcoin. Since the shutdown of Silk Road, articles in mainstream media have taken a more serious look at how Bitcoin can be used. I also briefly attempted to mine bitcoins, but did not have hardware that was powerful enough to yield any bitcoins.

I found that throughout this study Bitcoin users reached out to me—many were interested in the results of the research. Some participants sent links to articles or images they thought I would find interesting. One Bitcoin user even gave me a generous gift on Reddit for appreciation of my survey—0.05 bitcoins. At the time this was valued at around $8 USD, but at the time of writing this paper, it is worth $19.71 USD. This gift mirrored traditional ethnographers' experiences of receiving gifts from members of the communities they research (Nardi, 2014).

## 5. Findings

My findings examine the reasons Bitcoin users turned to algorithmic authority, and the ways in which that authority was mediated by human judgment. My findings reveal some of the tensions and complexities of algorithmic authority. Participants demonstrated the difficulty in determining whether an algorithm is political or apolitical, centralized or decentralized, promoting resistance or reifying institutional hegemony. These binaries were further complicated by the difference in how Bitcoin users wanted the currency to function and the ways in which it was actually used. The

following is an overview of the different tensions that I observed:

- *Political or apolitical:* Some participants saw Bitcoin's algorithms as the sorts of apolitical and incorruptible tools that Gillespie believed algorithms tend to be presented as being. As a result, they preferred them over the practices of existing institutions which were governed by corruptible human beings. However, some participants saw this departure from traditional institutions as a political choice. For some participants, it was meaningful to distinguish whether they felt that the project was political or apolitical because the distinction aligned them with certain viewpoints on Bitcoin's purpose.

- *Centralized or decentralized:* Despite the consensus among the participants that human lead institutions are less trustworthy than those run by algorithms, some participants also argued that some level of human oversight and judgment were necessary for Bitcoin to function smoothly. These participants were not necessarily arguing for hierarchical governance, but rather through consensus on major decisions and a neo-liberal sense of personal accountability for decisions that affected a small number of people. The role of human intervention was debated among Bitcoin users and some felt that certain methods of intervention were contradictory to Bitcoin's decentralized values.

- *Promoting resistance or reifying institutional hegemony:* Many participants argued that existing institutions—the very ones that they rejected—needed to support Bitcoin for it to gain widespread adoption. While a number of participants indicated that widespread adoption would ultimately be positive for Bitcoin, some were concerned that this adoption would cause Bitcoin to become like the very institutions they opposed. For participants who raised these concerns, the authority of cryptographic algorithms was decoupled from the authority of Bitcoin, and they said that they would start using another crypto-currency if Bitcoin became centralized.

- *Contradictions between desires and reality:* Participants held many utopian ideals and hopes for the ways that Bitcoin could change the world. However, many acknowledged that the social norms and regulations around Bitcoin are still being developed. It was often compared to the Wild West or the early days of the internet, and while they believed its transformative power for economic systems could be as great as the internet was for telecommunication systems, it was unclear to some participants when and how that would happen. Many participants admitted that they did not even use Bitcoin as a currency because they had nowhere to spend it. For many participants, Bitcoin was more like a stock or a commodity.

Exploring these tensions allows us to gain a greater understanding of the complexities of algorithmic authority, the different multiplicity of ways in which algorithmic authority may manifest and be understood, and how this authority is continually renegotiated.

### 5.1. Demographics of Bitcoin

The Bitcoin users who responded to my survey were predominantly American (51%), male (96%), libertarian (60%), and between 25-34 years of age (50%). 63% had bachelor or graduate degrees. Participants were relatively affluent; only 27% self reported making less money than the average person in their country. While gathering demographic data was not the main focus of this study, it has been included here in order to give readers a better sense of the background of participants and how that background might inform their views.

When I shared some preliminary results with the /r/bitcoin community, I was asked by users if the demographics of Reddit might be skewing the data. A reddit user responded with, "*Everything about this confirms people's beliefs about /r/Bitcoin. And the majority demographic being 25 and 34 year old, heterosexual, atheist or agnostic, American, and male is extremely reflective of reddit in general. I suppose an interesting question would be if this study is reflective of all Bitcoin users. I'm pretty sure it is.*"[10] I did find that participant demographics were similar to those of Reddit as a whole. According to a

---

10 http://www.reddit.com/r/Bitcoin/comments/1yvej4/follow_up_my_research_survey_on_bitcoin/cfo4794

2011 survey[11], Reddit users tended to be male (78%), young (84% were between 18 and 34), and American (64%). However, it is interesting to note that the Bitcoin community had a significantly higher proportion of males, was more geographically dispersed, and a bit older than Reddit users.

**Table 1: Demographics of Bitcoin**

| Demographic Category | Percentage |
|---|---|
| *Gender (allowed to pick multiple options)* | |
| Male | 96.92% |
| Female | 2.09% |
| Other | 1.86% |
| *Self-reported income* | |
| Higher than national average of participant's | 44.71% |
| Around the national average of the | 27.04% |
| Less than the national average of the participant's country | 27.88% |
| *Education* | |
| Less than a high school degree | 2.11% |
| High school degree or equivalent | 7.28% |
| Some college but no degree | 23.24% |
| Associate degree | 4.69% |
| Bachelor degree | 35.92% |
| Graduate degree | 26.76% |
| *Age* | |
| 18 to 24 | 18.82% |
| 25 to 34 | 50.12% |
| 35 to 44 | 21.41% |
| 45 to 54 | 6.82% |
| 55 to 64 | 2.35% |
| 65 to 74 | 0.24% |
| 75 or older | 0.24% |
| *Political beliefs (allowed to select multiple* | |
| Libertarian | 59.25% |
| Moderate | 36.25% |
| Anarchist | 27% |
| Left-wing | 25.25% |
| Green | 18% |
| Socialist | 11% |
| Right-wing | 8.25% |
| Communist | 2.5% |

The preponderance of males in my survey was not a fluke—in 2014, a survey of Bitcoin users found that 93% of respondents were male[12]. This was a slight improvement on the gender imbalance seen in a 2013 survey which had found that 95% of the respondents were male[13]. Speculation about why Bitcoin is overwhelmingly used by men is beyond the scope of this paper[14].

---

[11] http://www.redditblog.com/2011/09/who-in-world-is-reddit-results-are-in.html
[12] http://simulacrum.cc/2014/02/01/bitcoin-community-survey-2014/
[13] http://simulacrum.cc/2013/03/04/the-demographics-of-bitcoin-part-1-updated/
[14] Two participants did relate stories about how they had tried to speak up in favor of feminist issues on /r/bitcoin and BitcoinTalk and received very negative reactions from other users.

Participants' political beliefs were varied. Many chose multiple political labels for themselves out of the eight that were provided, i.e., anarchist, communist, green, left-wing, libertarian, moderate, right-wing, and socialist. 53% of participants selected more than one label for an average of 1.87 labels per participant. For those that selected at least one political label, 59% selected libertarian. However, an open-ended question about political beliefs showed that many participants had differing opinions about what these labels meant.

Survey participants came from 48 countries. The countries with more than 1% of the participants included: US (51%), Germany (7%), UK (6%), Canada (6%), Australia (4%), Netherlands (2%), Sweden (1%), Finland (1%), Norway (1%). There were 84 participants in 39 other countries, each with less than 1% of the participants: Switzerland, France, Singapore, Russian Federation, Poland, Belgium, Czech Republic, New Zealand, Spain, Italy, India, Ireland, Austria, Slovenia, Greece, Philippines, Argentina, Romania, Denmark, Croatia, China, Serbia, Israel, Brazil, Portugal, Japan, South Korea, Belarus, Malaysia, Slovakia, Mexico, South Africa, Moldova, Hungary, Lithuania, Taiwan, Thailand, Ukraine, and Benin.

Of the 22 interview participants, 13 were from the US, two from Australia, two from Germany, one from Argentina, one from Canada, one from Croatia, and one from India. One was an American expatriate living in China.

### 5.2. Algorithms are more trustworthy and authoritative than existing institutions

In this section I examine the reasons algorithmic authority was preferred over the authority of existing institutions such as governments or banks. I argue that for many, using Bitcoin was an act of resistance against institutions they felt had failed them. Beer (2009) argued that "algorithms are carving out new complex digital divides that emerge in unforeseen and often unnoticed ways in the lives of individual agents" and that it will be difficult to identify and research the ways in which people resist these algorithms. However, in a reversal of Beer's concern, in this section I explore the ways in which *algorithms can explicitly and visibly act as resistance* to institutions. This is not to say that Bitcoin is an inherently resistive technology, but that it has the capacity to be used in that way.

Some Bitcoin users were drawn to the currency because of their dissatisfaction with current economic practices, particularly with how governments can print money at will, causing inflation. One survey participant explained, "*Since we couldn't elect officials to be fiscally responsible and reign in the Federal Reserve, I actually have the freedom now, and means to preserve my savings and wealth through a non-inflationary currency. Everything else attractive about bitcoin is a bonus.*" Another survey participant stated that he liked Bitcoin because it "*is the convergence of technology (open source, p2p, cryptography) that is really going to change the world for the better and the more people that know about it the sooner we can get away from a debt based inflation run economy.*"

For participants who felt disenfranchised by governments and banks, Bitcoin offered an alternative. One participant, Terry, spoke about his distaste for banks. He had worked at a bank for years, and felt that banks did not act in the best interest of the people: "*I felt like I could punish the bank because I was able to write down finance charges and late fees and I was able to change interest rates*." Tom, a participant who was particularly concerned with government corruption, argued that fiat currency is coercive and violent: "*People can at last choose a form of money that isn't controlled by an entity which will shoot you if you misuse it.*" These participants lived in places such as the United States and Canada where they felt that the infrastructure was functional, but did not align with their morals. For these participants, Bitcoin was financial freedom from the forces governing fiat currency, which they felt that they had little to no control over, even in the case of Terry, who worked for a bank. Bitcoin, on the other hand, despite all of its technical complexity, gave the participants the feeling of being in control because of its transparency.

For some participants, Bitcoin was not just a technology of resistance, but also fulfilled practical needs. Franco, a participant in Argentina, explained that in his country, the official exchange rate between US dollars and Argentinian pesos was much worse than the black market exchange rate. He was employed by an American company and asked to be paid in bitcoins rather than US dollars in order to avoid dealing with either type of exchange. Roy, an American living in China, explained that his use for Bitcoin was based on getting around government restrictions: "*One of the best uses I've found for it is that it's the easiest way for me to get money from China into my American bank account. China has strict capital controls and foreigners can only send something like $500 USD out of the country per day*." For these participants, Bitcoin was not just a method of resistance, but also a way to cope with institutions that were unable to meet practical needs.

Some participants viewed Bitcoin as more trustworthy than governments because they considered Bitcoin an apolitical project. They considered the algorithms that govern Bitcoin incorruptible and impartial. Tom said, "*The cool part is, that the functionality of the software isn't political. It's concrete. Therefore, accurate information corroborated by the code will eventually filter out to the public at large. You can't politicize a lie based on verifiable facts forever. All you can do is temporarily create a lie meme that propagates, and take advantage of it before it gets destroyed by the rational types.*" Tom was referring to how although malicious users could spread misinformation for their own gain, Bitcoin is governed by an algorithm that, with time, would allow the misinformation to be countered by the authority of the code. Ken, a participant who dabbled in Bitcoin microfinance voiced a similar opinion, "*bitcoin is pure capitalism, plain and simple. It's property (thanks IRS) you can exchange for other property, without asking anyone's permission or paying anyone but the people who keep the network running (miners). Capitalism exists in every type of political environment, whether open or underground. [...] Anarcho-capitalism is a politcal philosophy, and it happens to be one that's very compatible with a currency like bitcoin, since no central authority is necessary to issue it [...] I consider it an apolitical currency, as is any cash.*" For Ken, Bitcoin is a capitalist project, but in his view, capitalism is separated from politics.

This notion that Bitcoin is apolitical seems to directly go against the idea that "artifacts have politics" (Winner, 1980). However, as Ken demonstrated, "apolitical" may refer to being non-partisan, or in the case of Tom, it may refer to being perceived as being grounded in a factual concreteness. Gillespie (2014) makes an argument similar to Tom's in his essay "The Relevance of Algorithms," when he says, "That we are now turning to algorithms to identify what we need to know is as momentous as having relied on credentialed experts, the scientific method, common sense, or the word of God". Algorithms are replacing institutions that are not as impartial or apolitical as many would like them to be. There is a utopian sense that software can be developed without political biases or prejudices. For participants who had this view, Bitcoin is a versatile technology that can support many different political viewpoints and for them, the focus of their interest on Bitcoin was on the technological aspects of Bitcoin.

For the participants who viewed Bitcoin as a political project (and also for many of the ones that did not), Bitcoin provided not only an alternative to institutions that they disliked, but also a possibility for overthrowing them. Some participants imagined a future in which Bitcoin would be the global currency, uniting people across the world, and ridding them of the tyranny of banks. For these participants, Bitcoin held much of the same appeal as the early internet did. They were excited to be early adopters of a technology that they believed could transform the world and to have a hand in shaping its future.

## 5.3. Algorithmic authority is mediated by human judgment

In this section, I explore the ways in which Bitcoin's decentralized algorithmic authority required trust in a number of human actors. Participants preferred algorithms to institutions, but they argued that Bitcoin itself and third party Bitcoin services require human oversight. Users used their own judgments to take precautions to prevent theft or falling victim to scams. These kinds of human mediation suggest that the judgment of individuals is a necessary supplement to algorithmic authority. When I use term "mediation", I refer to Activity Theory's definition of mediation. Mediation is the ways in which humans use technology as a tool to act out their desires and intentions in the world (Kaptelinin and Nardi, 2006). However, algorithmic authority reverses this concept and uses human judgment as a tool for algorithms to act on the world. In this section, I will explain the main ways in which human judgment supplemented Bitcoin's algorithms.

In the absence of any formal centralized human authority, I found that Bitcoin users have to spend time and effort to discern which instances of human authority are legitimate. Most of the interview participants reported frequently spending 2-3 hours a day reading up on Bitcoin or communicating with other Bitcoin users. Jonathan was a participant who no longer felt that he had time to spend on Bitcoin, which perplexed me. What does it mean to spend time on a currency? When asked, he explained that Bitcoin could be time consuming because, "*First, it's complex and takes a while to understand. Then it is the constant revolutionary language—everything is about to change in an instant, so you keep checking in to witness that instant. You are ( or one is) constantly waiting for your small holdings to make you rich.*" While many participants explained that they enjoyed keeping up with a technology that changed so rapidly and enjoyed communicating with other Bitcoin users, they also needed to be informed so that they could learn who to trust, how to protect their bitcoins from theft or fraud, and what community interventions were necessary to help Bitcoin itself run smoothly.

A notable example of when human mediation would have prevented a disaster occurred in February, 2014. The largest Bitcoin exchange market, MtGox, shut down suddenly and filed for bankruptcy in Japan. Bitcoin users who had been storing money on the exchange's server suddenly lost their bitcoins. The closure was particularly notable because MtGox, founded in 2010, was handling 70% of *all* Bitcoin transactions by 2013 (Vigna, 2014). The owners of MtGox claimed that $480 million USD of bitcoins had gone missing (Dougherty and Huang, 2014).

Participants argued that MtGox should have had more organizational oversight. Frankenmint, a participant whose sole source of income was Bitcoin mining and investing, directly addressed the issue of human mediation, commenting: "*I think that Mark [the CEO of MtGox] did have a leak of coins in the Gox system, and refused to put the expertise together to have the exchange algorithm better regulated by humans. He blamed malleability instead of his systems which lacked human verification. His organization was beyond incompetent, in my opinion*." Ken, who had a large mining operation and was an active user of alternative crypto-currencies, stated, "*I also feel bad for the people who trusted Gox and didn't understand the implications of that trust.*" Earlier in this paper, I had defined trust as the ability to reasonably predict the actions of other actors. While MtGox did have codified rules of operation, a number of participants stressed that the world of Bitcoin is quickly evolving—bugs in the software are found, businesses shut down quickly or turn out to be scams, and regulations change.

While MtGox required organizational oversight, participants also considered individual judgment to be important. Most had at some point been scammed out of bitcoins or had made an investment that was not successful. Participants observed that with enough experience and time, it was not difficult to tell which services were trustworthy and which were scams. One participant said, "*If you want to know the most recent news about Bitcoin you have to spend a lot of time on it, but for the average consumer that's not really important.*" Users learned to look for services that were transparent and kept users informed. Frankenmint stated, "*I believe honesty and integrity are the most important as the community demands trust. There have been too many failed ventures and screw-ups, Mt Gox, simply being the largest quandary so far….2nd, having knowledge and being willing to share it with others [makes one perceived as more trustworthy].*"

Another area for human intervention was in preventing theft or fraud. Theft and fraud are unfortunately serious issues in the Bitcoin community because the transactions are irreversible and pseudo-anonymous. As a result, reputation is extremely important in the Bitcoin community, which means that in practice, Bitcoin users cannot always anonymous. One participant said that he had contributed to a website that a microfinance site that took all loans in bitcoins. He tried to only loan to people whom he thought would repay him, but he did not have great success. Tom had pre-purchased Bitcoin miners from Butterfly Labs, a company whose mining hardware was wildly popular among Bitcoin miners and thus it seemed legitimate, but he, and countless others, found that his order never materialized.

Participants argued that new Bitcoin users had difficulty understanding how to keep their bitcoins safe, and that users would have to "relearn" how to protect their money. Lawrence, a participant who was quite active on Reddit, hoped that "*[a]s more people get involved, they will learn quickly what is necessary to secure their bitcoins. Hopefully these same practices will carry over to other services, which can bring exposure on proper security to other users, therefore making it less intimidating when/if they checkout bitcoin.*" Participants viewed this knowledge as something that had been culturally lost in an age in which transactions can easily be reversed and credit card companies warn customers if their algorithms detect that the customer's identity has been stolen.

It was not just Bitcoin-related software that needed human judgment and intervention—Bitcoin users have had to step in to prevent Bitcoin itself from facing serious problems. When version 0.8 of the beta Bitcoin client software was released, most *miners* upgraded to the latest version, but most *users* did not. Because of a change in the code, the 0.8 software recognized a block of transactions that the 0.7 software did not. This discrepancy caused the two different versions of the software client to use different chains of transactions. A prominent member of the Bitcoin community, Vitalik Buterin, noted that to make sure that everyone used the same blockchain, the "mining pool operators came together on IRC chat" and decided that they had to intentionally cause a 51% attack in order to resolve the fork (2014). A 51% attack is launched when the entity with a majority of computing power in the Bitcoin network chooses to manipulate the blockchain for their own purposes. Buterin pointed out that this "attack […] was seen by the community as legitimate." According to Buterin's argument, Bitcoin users saw this temporary assemblage of centralized power as more legitimately authoritative than Bitcoin's algorithms, which they had to "attack".

Some participants saw centralization of human resources as a necessary evil. For example, some participants felt that the Bitcoin Foundation was necessary for Bitcoin to be taken seriously. The Bitcoin Foundation is the unofficial public face of Bitcoin in the United States, and interfaces with the American government to help shape the laws that govern Bitcoin. One survey participant said, "*Don't really like [the Bitcoin Foundation], but it is good to have some "legitimate" group trying to advance bitcoin interest in the political sphere. Their actual influence on bitcoin is pretty limited so that can't do much harm.*" Other participants disliked the Bitcoin Foundation because they felt that they did not represent their interests and only reflected American concerns about government regulation of Bitcoin.

### 5.4. Algorithms need institutional support

Participants argued that the greatest barrier to the use and adoption of Bitcoin was lack of third party support. Many participants did not actively use Bitcoin, and for those that did, it was largely as a symbolic gesture in support of Bitcoin. To gain wider support, participants felt that Bitcoin must be seen as a legitimate and reputable currency. As I noted earlier, Weber stated that legitimacy is a requirement for authority. While participants felt that Bitcoin was legitimate, they found that they sometimes had trouble convincing others of its legitimacy. Bitcoin may have rational-legal authority (and for some people, it may have charismatic authority if they find Satoshi a compelling figure), but it does not have traditional authority, which may be a major drawback for individuals who are more traditional themselves. One participant told me that he tried to give away small quantities of bitcoin to people who he thought might find it compelling and they almost always refused him.

This issue of legitimacy and authority is important because without it, participants were limited in what they could do with Bitcoin due to poor institutional support for the currency. Participants reported that brick-and-mortar stores that accepted bitcoin were few and far between. While some participants made purchases with Bitcoins at brick and mortar stores, such as restaurants and gas stations, most made their purchases online. Of the participants who went to brick and mortar stores, most reported some difficulty in completing the transaction. Franco told a story about how he tried to pay for a meal at a restaurant that proudly advertised that they accepted Bitcoins, but none of the employees knew how to charge him. They ended up calling the owner of the restaurant and walking through the transaction over the phone.

One method of using bitcoin at brick-and-mortar stores that did not accept the currency, was to use a website such as gyft.com to buy gift cards. Gyft.com uses the service BitPay to accept Bitcoin transactions and automatically transfer the bitcoins into local fiat currency. These methods of paying for goods with Bitcoins show that participants were unable to avoid using fiat currency completely. This was not due to hypocrisy but rather the difficulty in doing so (see (Hill, 2014) and (Craig & Craig, 2014) for examples of people have tried to live strictly on Bitcoin). This peculiar transformation of fiat currency to bitcoin in order to turn it back into fiat currency again exemplifies the complicated relationship between Bitcoin and fiat currency. For some Bitcoin users who purchased gift cards, it was for the novelty of the experience—many participants described Bitcoin as "fun"—but for others, it was a show of support for the currency.

As a result of this poor institutional support, many Bitcoin users simply amassed Bitcoins. One participant, Misal, wrote about the complexities of using Bitcoin in India. What Misal observed in India corresponded with what I heard from participants in other regions of the world as well. He said, *"In the Indian ecosystem, bitcoin is still just a holding asset – this is not going to help the Indian bitcoin community to grow. In India If people wants to make it a sustainable currency in the longer run, then there should be an intrinsic value to it. And that comes when people start using that currency."* Some participants who held onto the currency viewed it like a stock—they "worked" for Bitcoin by mining, and they were paid in company stock. It was in their best interest to promote their company, because it would increase the value of their stock. This behavior demonstrates the difference in how some users used bitcoin (as something other than a currency) and the stated values of most of the participants.

As noted earlier, many participants reported spending multiple hours a day on Bitcoin; however, it was uncommon to find participants who actually made frequent transactions. As of 2012, 78% of all Bitcoins mined were presumed to be out of circulation—accumulated at "addresses which only receive and never send any BTC's" (Ron and Shamir, 2012). This suggests that only 22% of bitcoins are actively in circulation. In a survey of Bitcoin conducted by a community member[15], 29% of respondents reported that they had never purchased anything with Bitcoins, and 62% had not purchased anything with bitcoins in the last 30 days. However, only 6% did not have any bitcoins, suggesting that most of the users who had not spent Bitcoins were in possession of them, but chose not to spend them. While some Bitcoins may be lost to people whose wallets they have encrypted and no longer remember the password for, people who no longer use Bitcoin, or people who treat Bitcoin as a long term investment, it was also the case that some Bitcoin users hold onto Bitcoins in the hopes that in the future there will be a better infrastructure in place for using Bitcoins.

In addition to Misal, there were participants who lamented the fact that it was not more useful in their regions as well. However, they did not feel that it was a problem with Bitcoin itself. Integrating Bitcoin into other institutions was necessary, either through laws or third-party applications. For this reason, one participant argued that regulation is *"[n]ecessary, unavoidable, and not so onerous as it has been made out to be. I think the major governments of the world recognized right off the bat that this couldn't be contained, so their moving to integrate it."*

A significant minority of participants (30% of survey participants) said that they were not necessarily opposed to some government regulation of Bitcoin. Participants gave two main reasons for this. First, they wanted Bitcoin to be recognized as a "legitimate" currency by mainstream society, which meant distancing Bitcoin from illegal activity. Second, they felt that regulation would make it easier for Bitcoin to be used with existing institutions. One survey participant said, *"Regulation is important in the financial sector. Bitcoins need to be able to be transacted without fear of criminal exploitation. This requires an empowered authority to prosecute fraudsters and other financial criminals. Anarchists will dispute any government intervention, but without established trust no market can succeed. Bitcoin cannot continue to be 'the Wild West currency' and also succeed in the long-term."*

Because Bitcoin allows for pseudo-anonymous transactions, it has been used as a currency of digital black markets, the most well known being the now defunct Silk Road, which was primarily a marketplace for drugs. However, most participants were adamantly against using Bitcoin for illegal purchases because they felt that the association would undermine Bitcoin's legitimacy and authority in the eyes of the greater public and governments. One survey participant said, "*People who have no idea what bitcoin is will be able to see that for example, Silk Road may be a bad thing (expressed by mainstream news sites), but then the Bitcoin Foundation pops up on their google search and maybe shows them that the bitcoin isn't just about drugs and illegal activities*." Another participant, Terry, said that although he had been interested in Bitcoin when he had first heard of it, he did not try it at first because he thought that it was only good for black market purchases and he did not want to be associated with that kind of activity.

---

[15] http://www.reddit.com/r/Bitcoin/comments/21a0n4/survey_results_bitcoiners_of_reddit/

Similarly, some participants were cautious about identifying with the libertarian label because they felt it might seem extreme. For example, Ken said, "*I consider myself a libertarian, but I think that word gets abused a bit*." Simon, a Bitcoin investor, had similar views: "*I think sometimes the very extreme libertarian perspectives will be detrimental to mainstream adoption and turn people off […] I've been finding myself more sympathetic to liberaterian views, but I prefer to keep a level head*". Lawrence expressed concern about how other users represented Bitcoin: "*I also like to think of myself as policing bitcoin's reputation. /r/Bitcoin is an important resource for people starting out or researching bitcoin. We do not need to come off as delusional.*"

A subset of users were concerned that the integration with existing institutions might fundamentally change what Bitcoin means in a broader social context and change how it is used. Keith, a participant who had written white papers on future uses of the blockchain and similar technologies, argued that Bitcoin will eventually evolve into the same centralized capitalist institution that many Bitcoin users oppose. He said, "*It only means we can have perhaps some time where it's decentralized until the arms race results in a sort of king/queen of mining […] it's similar to what happens with capitalism where you end up with big businesses, then mega business, then just a few businesses who control everything. Bitcoin will develop in a similar way until a few businesses control every aspect of it. So it's about always innovating and always having different altcoins [other crypto-currencies] in competition.*" Another participant, Colin, who ended up deciding to stop using and mining bitcoins three months after our interview, asked, "*How does the individual miner compete with the corporate mining farms? – well, they don't, do they?*"

For many participants, it was not Bitcoin that they had high hopes for, but the blockchain algorithm that Bitcoin employed. They stated that Bitcoin might not exist in the future, but they believed that crypto-currencies would endure in the future. One survey participant argued, "*Even if Bitcoin isn't 'the one', it—or whatever comes after it—will change how we use/what we think about money forever*." Alternative crypto-currencies were a divisive issue in the Bitcoin community; many participants felt that altcoins were unoriginal clones of Bitcoin that diluted the authority of Bitcoin. For these participants, the authority lay with Bitcoin itself. But for participants who feared that Bitcoin would become subsumed by a culture they rejected, alternatives were essential, and the blockchain algorithm itself was what held authority.

## 6. Discussion

I found considerable variance in how participants viewed the currency. Bitcoin users are a diverse bunch: one participant who imagined a future in which crypto-currencies were used as socialist tools to generate money so that everyone would have a base income to live off of, whereas another participant viewed Bitcoin as a tool of rightwing libertarian politics inspired by politicians such as Ron Paul. Some participants were crypto-anarchists. Some participants believed in radical transparency. There were participants who found the pseudo-anonymity of Bitcoin to be very important and other participants who disliked the anonymous aspect of Bitcoin. There were participants who wanted more government regulation of Bitcoin and others who hated the idea of any government regulation. While I did find that participants tended to fit certain demographics (e.g. participants were overwhelmingly male), participants clearly had varying political views. Most held beliefs that could broadly be described as libertarian, but those beliefs were articulated in different ways among participants.

In this paper, it is not my goal to state which views of the participants were "correct", but rather to demonstrate the ways in which an algorithm can still have authority over a diversity of users who interpret the purpose and functionality of the software in a multiplicity of ways. In this section, I will examine the ways in which algorithms gain authority over such a group of people and expand on my earlier discussion of algorithms, authority, and trust.

### 6.1. Authority and utopia

My participants were on a spectrum between two poles: those that hoped that Bitcoin could provide a new disruptive authority and those that felt that

Bitcoin is only valuable to the extent that it is utilized and integrated into existing institutions. For participants who were concerned with Bitcoin as a disruptive technology, the distribution of algorithmic authority across different socio-technical actors was problematic. For them, the appeal of Bitcoin was based on utopian visions of a technology unhindered by centralized institutions or human judgment. For these users, the blockchain algorithm was at the heart of the disruption; Bitcoin itself was just one application of the blockchain. The utopian vision that inspired Bitcoin is an example of an *incomplete utopian project* (Gregory, 2000). Gregory argues that utopian projects are characterized by persistence: "utopian projects outlive any particular attempt at realization, nor is any particular failure sufficient to spell the end of a utopian quest." The diversity of alternate crypto-currencies based on the blockchain algorithm provides the heterogeneity needed to continue with the utopian visions of Bitcoin users if Bitcoin fails to live up to their expectations.

And indeed, the growing centralization of Bitcoin's governance points to the impossibility of ever living up to the utopian ideals of the Bitcoin community. While it is true that some participants were in favor of some forms of regulation, they largely viewed it as a necessary evil and not as a desirable future for Bitcoin. Similarly to how participants had different definitions of what it means to be (a)political, Bitcoin users may understand centralization in different ways. For example, one Reddit user wrote about the issue with the 0.8 code (discussed in section 5.3.), "centralization was kind of important in mitigating this issue. we had big pools of miners able to do as needed instead of individuals. we also had a dev team we depended on to guide us through."[16] Another user replied: "It only needed centralization of leadership, rather than centralization of authority. We didn't need a central authority mandating a certain change. We just needed a central authority suggesting a change that others can choose to heed or ignore." For this person, centralization tended to imply a coercive form of authority that was undesirable[17].

While it is true that the algorithms of Bitcoin operate in decentralized ways, many other aspects of Bitcoin are not as decentralized as they seem at first glance. The failure of Bitcoin to be fully decentralized can be seen in internal regulation such as the Bitcoin Foundation, by governments, and by temporary configurations of centralized power such as the group of miners who decided to fix the issue with version 0.8 of the Bitcoin code. Furthermore, the code production may be open source, but the code is still produced in a hierarchical way. The Bitcoin code is free and open-source, however there are few programmers who currently work on the code—as of October 14, 2014 73.6% of the code was committed by only eight developers out of 252 contributors[18]. This sort of hierarchy is not uncommon in open source software:

> "Commons-based peer production, observe Benkler and Nissenbaum, emerges in environments driven by collaborative efforts and results from the meeting of free individuals allergic to 'managerial hierarchies'; but, as often happens with human things, the shattering of old hierarchies ends up producing new ones, as blatantly revealed by the statistical measures of online activities and by their compliance with the '80/20 rule' of power law distributions." (Miconi, 2013)

(And indeed—I calculated that 20% of the Bitcoin contributors *are* committing exactly 80% of the code[18].) It may be that Bitcoin, and other algorithms that are seen as "resistive" may only have authority as long as they are seen as such. As soon as they replace the old ways of doing things, they are no longer disruptive technologies, but instead are the new technologies to be disrupted. In Weberian terms, these algorithms may no longer be examples of charismatic or rational-legal authority, but become examples of traditional authority, which are in their nature conservative and resistant to change.

Keith Hart (2014) spoke to the ways in which Bitcoin-related services are much more traditional than they first appear: "Bitcoin—like neoliberalism that it mimics, dreams of markets and money without politics or the state. This dream is an illusion. If you don't accept that markets and money depend on politics, then the politics go underground. In the case of Bitcoin, 80% of all transactions were taking place with MtGox. It operated control over Bitcoin in a way that any central bank would desire, but could not realize."

For a diverse group of users, Bitcoin may be appealing because of its incompleteness, and, as Barocas et al. argued, the difficulty of understanding algorithms makes it easy to attribute all kinds of values and effects to them. Bitcoin may have authority simply because the values in the code can be interpreted in so many different ways by users. However, as Hart points out, in practice Bitcoin may be more traditional than its users would desire. However, my research suggests that this mismatch was not enough for participants to abandon the currency or their utopian ideals. While I cannot generalize about Bitcoin users as whole, I did find that 55.2% of survey participants had been using Bitcoin for over a year, which suggests that some bitcoin users stay active in the community long after becoming familiar with the ways in which Bitcoin may not adhere to their utopian ideals.

---

[16] http://www.reddit.com/r/Bitcoin/comments/1a4ab0/alert_payattention_a_block_was_mined_that_was_too/c8u0j3j

[17] Weber distinguishes authority from power by arguing that authority is supported by legitimacy whereas power is supported by coercion. If we use Weber's terminology, then the second commenter was arguing that centralization is not always coercive, but can be legitimately authoritative.

[18] Calculated by using bitcoin.org/development

## 6.2. Nuance and trust

At the beginning of the paper, I argued that trust in an actor comes from being able to predict how that actor will behave, something that is particularly easy to predict when users have an understanding of the open-source code of a project like Bitcoin, but much more difficult to predict when it comes to opaque, large institutions.

This notion of trust in Bitcoin's code has been expanded on by Andreas Antonopoulos, the Chief Security Officer of blockchain.info, in his article, "Bitcoin Security Model: Trust in Code" (2014). He says that the most important feature of this new model of trust is that, "[n]o one actor is trusted, and no one needs to be trusted. […] Trust does not depend on excluding bad actors, as they cannot 'fake' trust. They cannot pretend to be the trusted party, as there is none." According to Antonopoulos, as long as over half of the computing power is controlled by what Nakamoto referred to as "honest nodes", the decentralized and aggregated computing power of the network can be trusted. Users do not need to trust any other individual user in order to trust in Bitcoin, but they do need to trust the network as a whole. Antonopoulos' concept of why users trust the Bitcoin network suggests that Bitcoin is a heteromated system in which it is *essential* for human actors to offer their computing power (through mining) to the system, in order to make the system trustworthy. But I also found that Bitcoin users contributed to the functioning of Bitcoin in other heteromated ways as well. Bitcoin users helped maintain the code, avert crises, and assess the trustworthiness of third party organizations..

A similar perspective on trust was offered by Maurer et al. who argued that, "Bitcoin is […] all about trust—about eliminating the need to trust governments and corporations and about learning to trust the Bitcoin algorithm instead" (2013). Maurer et al. do acknowledge the power of the network in regulating code, but emphasize that users ultimately trust the underlying code that manages the network. This distinction between trusting the code and trusting the network that uses the code seems to be somewhat artificial—we could (and do) go one step further and say that users trust the algorithms that support the code. The distinction between network and code is an important one, though, in that one implies a trust in the wisdom of the masses, and another implies trust in technological processes.

I found that Bitcoin users had a more nuanced view of this trust than either of these theories might suggest; they recognized that it is not enough to just trust in the code or in the network. Bitcoin's code is subject to change and on rare occasions it has had serious errors (such as the issue with the 0.8 version of the software). Problems in the code can prevent the network from behaving correctly. Furthermore, although users may have placed authority in the code, as Maurer et al. suggested, or in the network, as Antonopoulos suggested, algorithms were unable to tell users whether to trust a specific vendor or whether to update their software right away.

It could be argued that trust in vendors is a different issue than trust in Bitcoin. However, without the involvement of third party vendors and services, Bitcoin does not have substance. Participants lamented the difficulties in using bitcoin for practical purposes—the most practical uses were for participants who needed it as an international means of exchange, but few used it for regular transactions. Therefore, trust in Bitcoin requires trust in more than just the code or network, but also trust in the ecosystem of services offered for the currency.

I argue that trust in algorithms refers to not just the algorithm itself but the uses of the algorithms. Although Bitcoin users say that they trust the code or that they trust the network, they actually go to great lengths to validate their trust in Bitcoin. Participants found that determining whether they could trust third party applications and services took an extraordinary amount of time because they could not defer judgment to institutions. For passionate users of Bitcoin, this extra effort to determine trust may not be a significant drawback, but for other users, Bitcoin may only become appealing once social norms and regulations create a centralized method for determining trust.

## 6.3. Bitcoin as Assemblage

As I have demonstrated, Bitcoin's algorithms gain authority not just through the actions of the code, but also by the actors who are necessary for the algorithm to run smoothly. Without the Bitcoin developers, the Bitcoin miners who process transactions, and the judgments of Bitcoin users, Bitcoin would cease to function. The same is true for other technologies that use algorithms to manage users—without the users there would be no algorithmic authority. Therefore, when we consider this emergent trend towards giving algorithms more authority, we must consider the relationships between the code, the users, and the institutions that support the application.

In my discussion of Amazon Mechanical Turk, I posited that *people confer authority on an algorithm when they feel that they can trust the assemblage of actors associated with the algorithm*. In the case of Bitcoin, developing this trust is difficult in the absence of formal centralization or rules that can prevent fraud or theft. Developing this trust without centralization is time consuming and may not scale well as both the Bitcoin user base and the number of Bitcoin related applications grow.

Returning to my questions at the beginning of the paper in the paper about whether algorithms can "do things" and what agency algorithms may have—I think these are the wrong questions to be asking. Not because they are unimportant, but because algorithms are so intertwined with other actors. It is clear from my study that there are a multitude of reasons that users turned to Bitcoin and a multitude of possible futures that users imagined for the currency. Even though many Bitcoin users profess to trust in the algorithms or the code, my participants found that it was also necessary to place some trust in the other actors in the assemblage.

## 7. Conclusion

In this paper, I examined the concept of algorithmic authority and discussed the ways in which Bitcoin users trust in the code and in their own judgment. I found that algorithmic authority does not just reside in code, but in a diversity of sociotechnical actors. However, it is still unclear how this authority should be distributed and utilized. I found that my participants were of two minds about the potential algorithmic authority of Bitcoin. One group hoped that Bitcoin would provide a new disruptive authority with its blockchain mechanism, and all that enabled for many possible applications. The other group felt that Bitcoin is only valuable to the extent that it is utilized and integrated into existing institutions. Participants concerned with Bitcoin as a disruptive technology saw the distribution of algorithmic authority across different sociotechnical actors as problematic. For them, the appeal of Bitcoin was based on utopian visions of a technology unhindered by centralized institutions. For these users, the blockchain algorithm was at the heart of the disruption; Bitcoin itself was just one application of the blockchain. The diversity of alternate crypto-currencies based on the blockchain algorithm provides the heterogeneity needed to continue promoting the utopian visions of Bitcoin users if Bitcoin fails to live up to their expectations.

In my research I identified two different ways in which algorithms could appeal to a large, diverse group of people: first, when values remain open ended, users can position themselves relative to the technology in a way that best suits their own values. Secondly, algorithms may be most desirable to users when they feel that they can also trust the developers, other users, regulators, and anyone else who may use or influence the algorithm.

More research is needed to better understand the ways in which algorithmic authority can best be used to empower users. As Bitcoin evolves, more research will also be needed to understand the relationship between the centralized institutions so many of its users oppose and the decentralized algorithmic authority of Bitcoin.